# Expected performance of the GAW Cherenkov Telewscope Array. Simulation and Analysis

M.C. Maccarone [1], P. Assis [2], O. Catalano [1], G. Cusumano [1], M.C. Espirito Santo[2],
P. Goncalves [2], M. Moles [3], M. Pimenta [2], A. Pina [2], B. Sacco [1], B. Tome' [2]

[1] Istituto di Astrofisica Spaziale e Fisica Cosmica di Palermo, IASF-Pa/INAF, Via Ugo La Malfa 153, 90146 Palermo, Italy
[2] Lab. Instrumentacao e Fìsica de Partìculas, LIP, Av. Elias Garcia 14, 1000-149, Lisbon, Portugal
[3] Instituto de Astrofìsica de Andalucìa, IAA, Camino Bajo de Huétor 50, 18008, Granada, Spain

**Abstract**
GAW is a "path-finder" experiment to test the feasibility of a new generation of Imaging Atmospheric Cherenkov telescopes that join high flux sensitivity with large field of view capability using Fresnel lens, stereoscopic observational approach, and single photon counting mode. GAW is an array of three telescopes that will be erected at the Calar Alto Observatory site (Spain, 2150 m a.s.l.). To evaluate the performance of GAW, a consistent dataset has been simulated, including a Crab-like source observation, and a proper image analysis code has been developed, as described in this contribution. The expected performance of GAW are also reported, mainly for what concerns effective area, angular resolution, Cherenkov flux as function of the core distance, ability in the gamma/proton separation, and sensitivity. The first telescope realization, foreseen within the end of this year, will allow to verify if the parameter used in the analysis are in agreement with the "real" performance of the GAW apparatus.

**Introduction**
The Imaging Atmospheric Cherenkov (IAC) technique is an advanced and consolidated method to detect Extensive Air Showers (EAS) by measuring the Cherenkov light produced by relativistic charged particles induced by primaries crossing the Earth atmosphere [1, 2]. This technique opened a window to the ground-based gamma-ray astronomy in the Very High Energy (VHE) range (from 50 GeV up to several TeV) and continuously improves in what concerns detection performance and sensitivity, using multiple telescopes in stereoscopy (H.E.S.S., VERITAS, CANGAROO III).
Basically, a classical IAC telescope consists of an optical system formed by a highly reflectivity mirror, and of a focal multi-pixel camera operating in charge integration mode and characterized by Field of View (FoV) of the order of few degrees. To improve the sensitivity at the lowest energies for such a kind of telescope, considerable efforts have been made, mainly designing systems with larger mirror area (up to 17 m diameter, MAGIC).
A different approach distinguishes GAW [3]: its optical system consists of a refractive Fresnel lens (2.13 m diameter) optimized for a FoV of 24°x24°, and its focal camera is a grid of multi-anode photomultipliers (MAPMT) operating in Single Photon Counting (SPC) mode [4]. GAW is formed by three identical telescopes, posed at the vertexes of a quasi-equilateral triangle (80 m side). The SPC mode, together with the stereoscopic observational approach proper of the current IAC technique, will guarantee a threshold of few hundreds of GeV in spite of the relatively small dimension of the lens. The array will be located at Calar Alto Observatory (Sierra de Los Filabres, Andalucía, Spain) at 2150 m a.s.l.; the first telescope is expected to be erected by the end of 2007.
Science objectives, general description and technicalities about GAW are given elsewhere in these proceedings [5, 6]. Here we present and discuss the expected performance of GAW resulting from an "ad hoc" end-to-end simulation chain that takes into account all the elements and features of our system.

**The end-to-end simulation chain**

The CORSIKA simulation code [7] (version 6.5, with QGSJET as high-energy hadronic interaction model) has been used to generate the Cherenkov light, at single photon level, associated to air showers induced by gamma or proton primaries in the energy range 0.3--30 TeV, and detected at 2150 m a.s.l.. The effects of the atmosphere and a set of detector parameters values were included; in particular: atmosphere transparency, optics efficiency and MAPMT quantum efficiency as a function of the Cherenkov photon wavelength, and a constant reduction factor to take into account light guide transparency[i] and photomultiplier collecting efficiency [3]. Each shower was simulated with core location randomly chosen inside a square of 800 m side. The Cherenkov light produced by the CORSIKA showers was further enriched by Poissonian Night Sky Background (NSB) diffuse light, uniformly distributed in the camera pixels [ii].

The second step of the simulation chain concerns the GAW electronics and triggering system [3, 4]. The SPC mode adopted in GAW does not keep memory of the number of photoelectrons converted by each pixel of the MAPMTs, and the shower image will simply appear as a binary image with "pixels on" (*pixon*) in locations of the focal camera. In each telescope, the triggering will be activated if it presents, inside a logical region of 2x2 MAPMTs, a number of *pixon* greater than a given threshold. As detailed elsewhere in these proceedings [5, 6], a threshold of 14 *pixon* per event has been adopted; such a choice reduces to a negligible value the fake trigger rate due to the diffuse NSB light (0.0008 pixon/pixel/ns) and allows to detect showers of few hundreds GeV. An example of Cherenkov light imaged by GAW is shown in Figure 1.

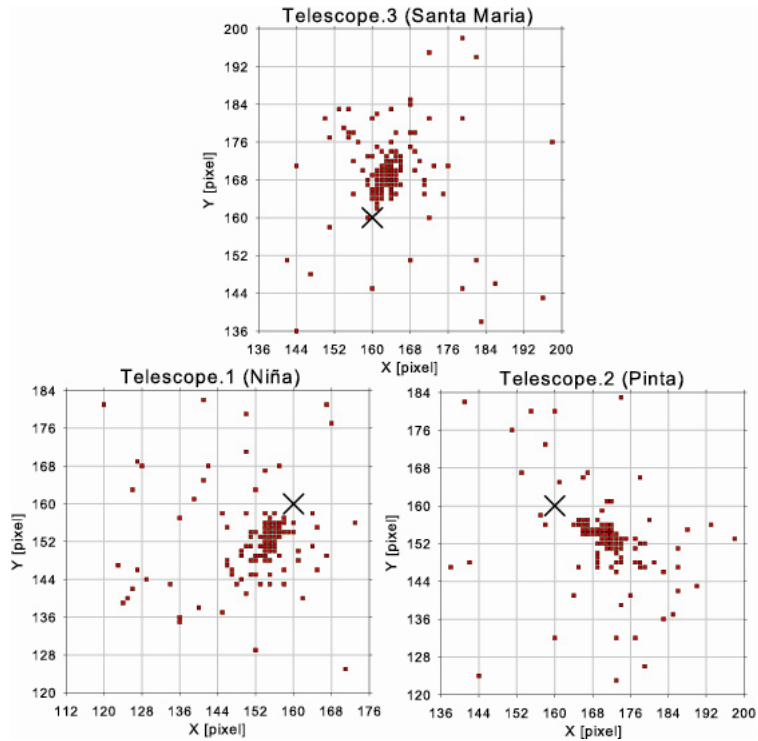

Figure 1: Example of a 2 TeV Cherenkov event as imaged by GAW (array layout not in scale). Only pixels of the most significant MAPMTs are shown. The big cross refers to the FoV center.

The simulation chain is completed by the image analysis step that makes use of clustering methods, probability theory and geometrical techniques [8]. The set of *pixon* forming the shower image appears as a rough ellipse, more or less spread and elongated, originated by the shower signature and embedded in the diffuse NSB background. To extract the relevant information, i.e. to separate the "good" signal against the background, we apply a robust single-link clustering algorithm [9] based on Euclidean metrics and Uniformity Test. A first cluster analysis step cleans the shower image by eliminating isolated (outliers) *pixon*; a second analysis step determines the shower image main axis through the quadratic moments of the cleaned image, by minimizing its distance respect to the location of the surviving *pixon* [3, 10]; all these outcomes allow us to describe each single shower image in terms of Hillas' parameters. Finally, thanks to the GAW stereoscopy, it is straightforward to determine the point that minimizes the distance from the main axes of the three Cherenkov images, and assume it as the shower arrival direction.

## GAW Expected Performance

### General performance

The results presented in this paragraph was obtained from the analysis of more than 400 000 gamma-induced showers (at least 20 000 for each energy) originated from a source at zenith angle of 0°, at random core location. Figure 2 shows the GAW effective collecting area ($m^2$) evaluated at the trigger level, and the detection rate (arbitrary units) of a Crab-like source. Under the constraint of trigger coincidence in all the three telescopes (3-fold) the Effective Area reaches the value of ~$6 \times 10^5$ $m^2$ at the highest energy. The convolution with the Crab spectrum shows a peak at 0.7 TeV, a positive result given the GAW relative small pupil dimension.

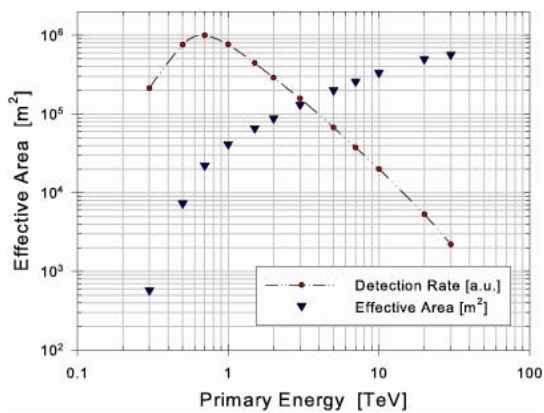

Figure 2: GAW Effective Area vs Energy at level of trigger and detection rate (arbitrary units) of a Crab-like source.

The number of *pixon* forming the shower image depends on the distance from the shower core and it allows the reconstruction of the energy of the primary. The distribution of the mean number of *pixon* inside a region of interest of ~5°x5° versus the core distance (after reduction in the proper telescope reference system) is reported in Fig. 3. A set of few selection criteria was adopted to continue the analysis, and related threshold values were defined; in particular, the number of surviving *pixon* must exceed the minimum (order of 10) required to apply image analysis methods. Under such a selection, we evaluated the angular distance, or Radius, between the reconstructed and simulated arrival direction of the shower, impinging within a core distance of 120m. The GAW Angular Resolution has been then defined as the value of Radius containing the 68% of reconstructed events; results are shown in Figure 4.

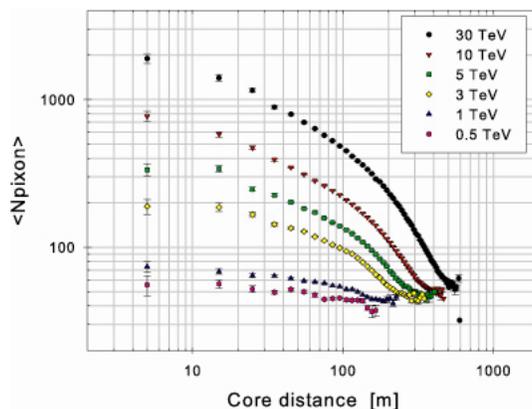

Figure 3: Distribution of the average number of *pixon* vs the radial distance from the shower core, 10 m binned, vs energy. Error bars reflect the standard error on the mean value estimation.

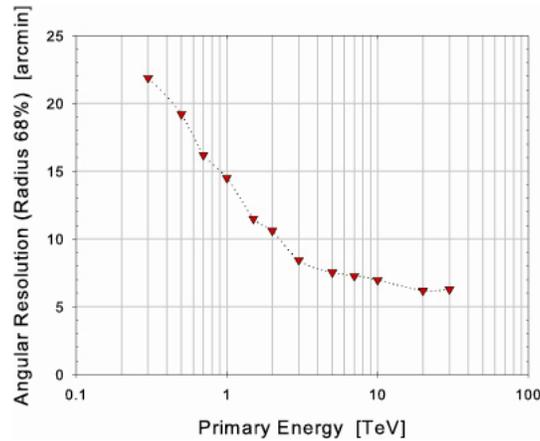

Figure 4: GAW Angular Resolution vs Energy for source on-axis. Core distance within 120 m.

## Simulating a Crab-like source observation

The sensitivity of GAW and its ability in the gamma/proton separation were evaluated simulating a pursuit observation of a Crab-like source. We have considered the path of Crab along the Calar Alto sky during a winter night; the simulated data-set (gamma and protons) refers to a period of 7 hours, subdivided in steps of 20 minutes, in the energy range 0.3--30 TeV. Each shower was simulated with core location randomly chosen inside a square of 800 m side. The GAW telescopes were maintained always on-axis with the gamma-induced shower axis. The proton-induced showers were simulated inside a 4° cone around the source position. The spectral distribution was defined with a slope of -2.49 and -2.74 for gamma- and proton-induced showers, respectively. Fig.5 shows the distribution, for gamma- and proton-induced showers, of the Hillas' parameter *AZwidth*[iii], averaged onto the 3 telescopes forming GAW. A proper selection cut on *AZwidth* will help us in the gamma/proton separation and then in the evaluation of the sensitivity of GAW, as reported in Table 1.

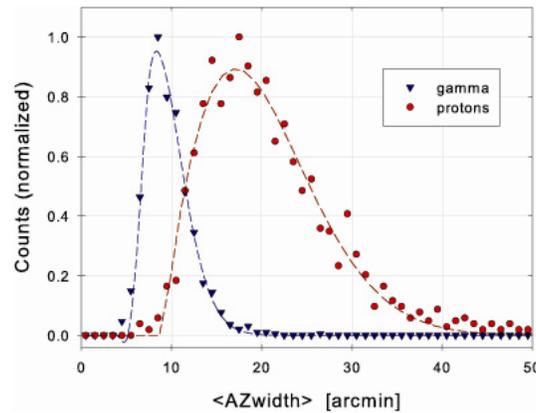

Figure 5: Distribution of the *AZwidth* Hillas' parameter for the simulated Crab-like source observation.

| | |
|---|---|
| AZwidth cut (60% -surviving) | 9.94 arcmin |
| Protons-surviving | 2.5% |
| Selection Cone, SC (68% -surviving) | 18. arcmin |
| S, number of gamma in 7 hour in SC | 237. |
| N, number of protons in 7 hour in SC | 20 |
| SNR per night (7 hours observation) | 15. |

Table 1: Evaluation of the GAW sensitivity simulating a Crab-like source observation in the energy range 0.3-30 TeV. The signal-to-noise ratio SNR is here defined as S/sqrt(S+N).

## Conclusions

The first telescope realization, foreseen within the end of this year, will allow of verifying if both the simulation and the parameters used in the analysis are in agreement with the "real" performance of the GAW apparatus. In fact, the simulation procedure here described, applied to only-one telescope, predicts that we can observe the Crab nebula in one night at least at 8-sigma detection level.

---

[i] To correct the dead area of the photomultipliers, each MAPMT pixel is coupled to a Light Guide which allows to uniformly cover the FoV.

[ii] 2200 photons m$^{-2}$ ns$^{-1}$ sr$^{-1}$, mean value of NSB as measured in the Calar Alto site [3], of the same order obtained with other similar measurements [11].

[iii] *AZwidth*, or *Azimuthal Width*, is defined as the Root Mean Square image width relative to the *Distance*' axis which joins the source to the centroid of the image.